\documentclass[11pt,a4paper]{article}

\pdfoutput=1

\usepackage{graphicx}
\usepackage{epstopdf}
\usepackage{jheppub}
\usepackage{hyperref}
\newcommand{\ee}{\end{equation}}
\newcommand{\br}{\begin{eqnarray}}
\newcommand{\bea}{\begin{eqnarray}}
\newcommand{\eea}{\end{eqnarray}}
\newcommand{\er}{\end{eqnarray}}
\newcommand{\ba}{\begin{array}}
\newcommand{\ea}{\end{array}}
\newcommand{\nn}{\nonumber}
\newcommand{\bi}{\begin{itemize}}
\newcommand{\ei}{\end{itemize}}
\newcommand{\bn}{\begin{enumerate}}
\newcommand{\en}{\end{enumerate}}
\newcommand{\bc}{\begin{center}}
\newcommand{\ec}{\end{center}}

\newcommand{\beq}{\begin{equation}}
\newcommand{\eeq}{\end{equation}}

\newcommand{\MP}{{M_{\rm P}}}
\newcommand{\arcsinh}{\text{arcsinh}}

\title{\centering Non-minimal (self-)running inflation:\\ metric vs. Palatini formulation}

\author[a]{Antonio Racioppi}

\affiliation[a]{National Institute of Chemical Physics and Biophysics, R\"avala 10, 10143 Tallinn, Estonia}

\emailAdd{Antonio.Racioppi@kbfi.ee}

\abstract{We consider a model of quartic inflation where the inflaton is coupled non-minimally to gravity and the self-induced radiative corrections to its effective potential are dominant. We perform a comparative analysis considering two different formulations of gravity, metric or Palatini, and two different choices for the renormalization scale, widely known as prescription I and II. Moreover we comment on the eventual compatibility of the results with the final data release of the Planck mission.}

\keywords{Inflation, non-minimal coupling, radiative corrections, Palatini formulation}

\begin{document}
\maketitle

\section{Introduction} \label{sec:Introduction}

According to the theory of cosmic inflation~\cite{Starobinsky:1980te,Guth:1980zm,Linde:1981mu,Albrecht:1982wi}, our Universe underwent a period of exponential expansion during the initial moments of its life. Inflation has the merit of providing at the same time a solution to issues like the flatness and horizon problems of the Universe and a way to generate primordial inhomogeneities, whose power spectrum is currently being tested in several experiments~\cite{Ade:2015tva,Ade:2015xua,Ade:2015lrj,Array:2015xqh,Planck2018:inflation}. In particular, the final data release of the Planck mission~\cite{Planck2018:inflation} casts strong constraints on the tensor-to-scalar ratio $r$, an observable related to the amplitude of primordial gravitational waves and to the scale of inflation.  As a consequence, the predictions of the simple monomial inflation models are ruled out at $2 \sigma$ level, leaving non-minimally coupled to gravity models as some of the most favorite ones. In this article we are going to study models of inflation with a non-minimal coupling to gravity of the type $\xi \phi^2 R$, where $\phi$ is the inflaton field, $R$ the Ricci scalar and $\xi$ a coupling constant. Similar models have been studied in a large number of works over the past decades (in e.g.\cite{Futamase:1987ua,Salopek:1988qh,Fakir:1990eg,Amendola:1990nn,Kaiser:1994vs,Bezrukov:2007ep,Park:2008hz,Linde:2011nh,Kaiser:2013sna,Kallosh:2013maa,Kallosh:2013daa,Kallosh:2013tua,Galante:2014ifa,Jarv:2016sow,Chiba:2014sva,Boubekeur:2015xza,Pieroni:2015cma,Salvio:2017xul,Bostan:2018evz,Almeida:2018pir,Cheng:2018axr,Tang:2018mhn,SravanKumar:2018tgk,Kubo:2018kho,Canko:2019mud,Okada:2019opp,Karam:2017rpw,Karam:2018mft}). 
These models are particular interesting, since non-minimal couplings should be interpreted as a generic ingredient of consistent model building, arising from quantum corrections in a curved space-time \cite{Birrell:1982ix}. In particular, this is the case for the scenario where the Standard Model Higgs scalar is the inflaton field \cite{Bezrukov:2007ep}.
Comparisons of non-minimally coupled models of chaotic inflation were performed in e.g.\ \cite{Linde:2011nh,Kaiser:2013sna,Kallosh:2013maa,Kallosh:2013daa,Kallosh:2013tua,Galante:2014ifa,Jarv:2016sow}. In Refs. \cite{Kaiser:2013sna,Kallosh:2013tua}, it was shown that for large values of the non-minimal coupling, all models, independently of the original scalar potential, asymptote to a universal attractor: the Starobinsky model \cite{Starobinsky:1980te}. 
However, the presence of non-minimal couplings to gravity requires a discussion about the gravitational degrees of freedom. In the usual metric formulation of gravity the independent variables are the metric and its first derivatives, while in the Palatini formulation the independent variables are the metric and the connection. Using the Einstein-Hilbert Lagrangian, the two formalisms predict the same equations of motion and therefore describe equivalent physical theories. However, with non-minimal couplings between gravity and matter, such equivalence is lost and the two formulations describe different gravity theories \cite{Bauer:2008zj} and lead to different phenomenological results, as recently investigated in e.g. \cite{Bauer:2010jg,Tamanini:2010uq,Tenkanen:2017jih,Jarv:2017azx,Rasanen:2018ihz,Carrilho:2018ffi,Almeida:2018oid,Takahashi:2018brt,Antoniadis:2018yfq,Tenkanen:2019jiq,Tenkanen:2019xzn,Tenkanen:2019wsd,Bostan:2019wsd,Gialamas:2019nly,Rasanen:2017ivk,Kannike:2018zwn,Enckell:2018kkc,Racioppi:2017spw,Racioppi:2018zoy,Antoniadis:2018ywb,Bostan:2019uvv,Markkanen:2017tun,Enckell:2018hmo,Kozak:2018vlp}.
In particular, the attractor behaviour of the so-called $\xi$ attractor models \cite{Kallosh:2013tua} is lost in the Palatini formulation \cite{Jarv:2017azx}.  It is important to remark that in \cite{Kallosh:2013tua,Jarv:2017azx} the role of quantum corrections is implicitily assumed to be subdominant. On the other side, it has been demonstrated that radiative corrections to inflationary potentials may play a relevant role \cite{Kannike:2014mia,Marzola:2015xbh,Marzola:2016xgb,Dimopoulos:2017xox}, dynamically generating the Planck scale \cite{Kannike:2015apa,Kannike:2015kda}, predicting super-heavy dark matter \cite{Farzinnia:2015fka,Kannike:2016jfs} and leading to linear inflation predictions when a non-minimal coupling to gravity is added \cite{Kannike:2015kda,Rinaldi:2015yoa,Barrie:2016rnv,Artymowski:2016dlz,Racioppi:2017spw,Racioppi:2018zoy}. 

At the present date, all the comparative studies between the metric and Palatini formulations either consider only a classical tree-level analysis \cite{Bauer:2008zj,Bauer:2010jg,Tamanini:2010uq,Tenkanen:2017jih,Jarv:2017azx,Rasanen:2018ihz,Carrilho:2018ffi,Almeida:2018oid,Takahashi:2018brt,Antoniadis:2018yfq,Tenkanen:2019jiq,Tenkanen:2019xzn,Tenkanen:2019wsd,Bostan:2019wsd,Gialamas:2019nly} or assume that the leading contribution to radiative corrections is coming from some other additional particle (inside the Standard Model \cite{Rasanen:2017ivk,Kannike:2018zwn,Enckell:2018kkc} or beyond it \cite{Racioppi:2017spw,Racioppi:2018zoy,Antoniadis:2018ywb,Bostan:2019uvv}) rather than the inflaton itself. 
While there have been several works about self-induced radiative corrections for a non-minimal inflaton in the metric case (e.g. \cite{Buchbinder:1992rb,Elizalde:1993ew,Inagaki:2015fva,George:2013iia}), such a topic is completely unexplored in the Palatini one.
Therefore, the aim of this work is to fill the gap in the literature and present for the first time a comparative analysis of the possible gravity formulation (metric or Palatini) in the context of non-minimal inflation when self-corrections are the dominant loop contribution.

The article is organized as follows.
In section~\ref{sec:Preliminaries}  we set the notation reintroducing the main concepts about running coupling constants and the effective potential. In section \ref{sec:Non_min_G} we discuss the gravitational sector and the main differences between the metric and the Palatini formulation of a gravity theory. In section \ref{sec:inflation} we present the comparative study of the inflationary predictions. We conclude in section~\ref{sec:Summary}.

\section{Model building and effective potential} \label{sec:Preliminaries}

Consider the following action for a scalar-tensor theory in the Jordan frame
\begin{equation}
S = \!\! \int \!\! d^4x \sqrt{-g}\left(-\frac{M_P^2}{2}f(\phi)R(\Gamma) + \frac{(\partial \phi)^2}{2}  - V_{\rm eff}(\phi) \right) ,
\label{eq:JframeL}
\end{equation}
where $M_P$ is the reduced Planck mass, $R$ is the Ricci scalar constructed from a connection $\Gamma$ and $V_{\rm eff}(\phi)$ is the effective potential of the inflaton scalar. In the following we will focus on one particular type of $f$ function: 
\begin{equation}
f(\phi)=1 + \xi \frac{\phi^2}{M_P^2} \, ,  
  \label{eq:f:H}
\end{equation}
which is the usual Higgs-inflation \cite{Bezrukov:2007ep} non-minimal coupling
where we relaxed the condition that the inflaton is the Higgs boson and allowed the possibility that inflation is driven by another scalar beyond the Standard Model particle content. The tree-level inflaton potential is 
\begin{equation}
  V(\phi) = \frac{1}{4} \lambda \phi^4 \, ,
  \label{eq:V:tree}
\end{equation}
however our focus is on the 1-loop\footnote{While cosmological perturbations are invariant under frame transformations (see for instance \cite{Prokopec:2013zya,Jarv:2016sow}), the equivalence of the Einstein and Jordan frames at the quantum level is still to be established. In the present article we therefore apply the following strategy: first we compute the effective potential in the Jordan frame, eq.~\eqref{eq:Veff}, and consequently we move to the Einstein frame for computing the slow-roll parameters. Given a scalar potential in the Jordan frame, the cosmological perturbations are then independent, in the slow-roll approximation, of the choice of the frame in which the inflationary observables are evaluated~\cite{Prokopec:2013zya,Jarv:2016sow}. For further discussions on frames equivalence and/or loop corrections in scalar-tensor theories we refer the reader to Refs.~\cite{Jarv:2014hma,Kuusk:2015dda,Kuusk:2016rso,Flanagan:2004bz,Catena:2006bd,Barvinsky:2008ia,DeSimone:2008ei,
Barvinsky:2009fy,Barvinsky:2009ii,Steinwachs:2011zs,Chiba:2013mha,George:2013iia,Postma:2014vaa,
Kamenshchik:2014waa,George:2015nza,Miao:2015oba,Buchbinder:1992rb,Elizalde:1993ee,Elizalde:1993ew,Elizalde:1994im,Inagaki:2015fva,Burns:2016ric,Fumagalli:2016lls,Artymowski:2016dlz,Fumagalli:2016sof,Bezrukov:2017dyv,Karam:2017zno,Narain:2017mtu,Ruf:2017xon,Markkanen:2017tun,Markkanen:2018bfx,Ohta:2017trn,Ferreira:2018itt,Karam:2018squ}.} improved effective inflaton potential. Assuming that only self-corrections are relevant during inflation, the improved potential\footnote{Given the present constraint on the amplitude of scalar perturbations (see eq. (\ref{cobe})), the running of the non-minimal coupling $\xi$ can be safely neglected in the computation as long as the pertubativity of the theory is ensured (e. g. \cite{Marzola:2016xgb} and refs therein). More details on this topic are given in Appendix \ref{appendix}. A careful reader might also notice that quadratic curvature invariants are radiatively generated as well. In this paper we work in the linear curvature approximation (e. g.  \cite{Buchbinder:1992rb,Elizalde:1993ee,Elizalde:1993ew,Elizalde:1994im,Inagaki:2015fva} and refs. therein), leaving for a forthcoming paper the study of the impact of higher order curvature terms.} is 
\begin{equation}
 V_{\rm eff}(\phi,\mu) = \frac{1}{4} \lambda_{\rm eff} (\phi, \mu) \phi ^4
  \label{eq:Veff}
\end{equation}
where the effective quartic coupling is
\begin{equation}
 \lambda_{\rm eff}(\phi,\mu) = \lambda (\mu ) + 
 \frac{9 \lambda (\mu )^2}{ 16 \pi ^2} \ln \left[\frac{3  \lambda (\mu ) \phi^2}{\mu^2}\right]  \, .
  \label{eq:l:eff}
\end{equation}
The second part of eq. (\ref{eq:l:eff}) is the contribution coming from the Coleman-Weinberg (CW) 1-loop correction \cite{Coleman:1973jx} to the effective potential, while the first one comes from the renormalization group equation (RGE) \cite{PhysRevD.2.1541,Symanzik1970} of the quartic coupling, whose solution is
\begin{equation}
 \lambda(\mu) = \frac{\lambda _0}{1-\frac{9 \lambda _0 \ln \left(\frac{\mu ^2}{\mu
   _0^2}\right)}{16 \pi ^2}} \, ,
  \label{eq:l:run}
\end{equation}
where $\lambda_0 = \lambda(\mu_0)$ is the boundary condition for the RGE. For convenience we choose $\mu_0=\MP$ and keep $\lambda_0$ as a free parameter. It is important to keep in mind that the solution (\ref{eq:l:eff}) is correct at the order $O(\lambda^2(\mu))$, therefore any $\mu$-dependence at higher order can be thrown away and the effective coupling can be safely truncated as
\begin{equation}
 \lambda_{\rm eff}(\phi,\mu) = \lambda (\mu ) + 
 \frac{9 \lambda (\mu )^2}{16 \pi ^2} \ln \left(\frac{3 \lambda_0 \phi ^2 }{\mu^2}\right)  \, .
  \label{eq:l:eff:better}
\end{equation}

The purpose of the RGE improved effective potential is to obtain a potential that, at a given perturbative order, is independent on the choice of $\mu$ (e.g. \cite{Neubert:2019mrz} and references therein). Therefore, in the regime of validity of the RGE, i.e. until $\lambda(\mu)$ is small enough, any choice of $\mu$ is equivalent and should not carry any physical meaning \cite{Neubert:2019mrz}. Any effect coming from the choice of $\mu$ should be due to the loss of validity of the 1-loop expansion in eq. (\ref{eq:l:eff:better}) and the need for a result at least at 2-loops. However there are two choices which are quite popular and eventually convenient, which are \cite{Bezrukov:2013fka,Allison:2013uaa,Hamada:2016onh,Bostan:2019fvk}
\beq
 \mu_{\rm I}^2 = \frac{3 \lambda_0 \phi^2}{f(\phi)} \label{eq:PI}
\ee
also known as prescription I and
\beq
 \mu_{\rm II}^2 = 3 \lambda_0 \phi^2  \label{eq:PII}
\ee
also known as prescription II. Prescription I \cite{Bezrukov:2012hx} is the choice motivated by the scale-invariant quantization in the Jordan frame, while prescription II \cite {DeSimone:2008ei,Barvinsky:2009fy,Barvinsky:2009ii} corresponds to the usual quantization in the Jordan frame and it is convenient because it cancels explicitly the CW part of (\ref{eq:l:eff:better}), moving all the loop correction into the running of the quartic coupling. For convenience later on we will use the following notation: $\lambda_{\rm eff}(\phi,\mu_{\rm I,II}) = \lambda_{\rm I,II} (\phi)$.

It is also useful to notice that in case of very small $\lambda_0$, the dependence on $\mu$ explicitly cancels away. Performing a Taylor expansion of eq. (\ref{eq:l:eff:better}) till the 2nd order in $\lambda_0$ we get the following approximated effective coupling
\begin{eqnarray}
 \lambda_{\rm app}(\phi) &\simeq&  \lambda_0 + 
 \frac{9 \lambda_0^2}{16 \pi ^2} \ln \left(\frac{\mu ^2}{\MP^2}\right) +
 \frac{9 \lambda_0^2}{16 \pi ^2} \ln \left(\frac{3 \lambda_0 \phi^2 }{\mu^2}\right) = \nonumber\\
 &=& \lambda_0 +
 \frac{9 \lambda_0^2}{16 \pi ^2} \ln \left(\frac{3 \lambda_0 \phi^2 }{\MP^2}\right)
  \, ,
  \label{eq:l:app}
\end{eqnarray}
and the dependence on $\mu$ is completely removed. We notice that such expression recalls the running quartic coupling used in \cite{Racioppi:2018zoy} with $\delta = \frac{9 \lambda_0}{8 \pi^2}$, therefore we expect that some of the results of \cite{Racioppi:2018zoy} will be valid also here.

Because of perturbativity of the theory, the inflationary predictions of such a potential, in absence of a non-minimal coupling to gravity, are pretty similar to the ones of the tree-level quartic potential and already ruled out by data \cite{Planck2018:inflation}. Such predictions are dramatically changed if a non-minimal coupling to gravity is added, as we do in Lagrangian (\ref{eq:JframeL}). However, a modification of gravity calls for a discussion of what theory of gravity we are going to consider. This will be shortly discussed in the following section.

\section{Non-minimal gravity} \label{sec:Non_min_G}

We discuss now the gravitational sector and its non-minimal coupling to the inflaton.
In the metric formulation the connection is determined uniquely in function of the metric tensor, i.e.\ it is the Levi-Civita connection $\bar{\Gamma}=\bar{\Gamma}(g^{\mu\nu})$
\begin{eqnarray}
\label{eq:LC}
\overline{\Gamma}^{\lambda}_{\alpha \beta} = \frac{1}{2} g^{\lambda \rho} \left( \partial_{\alpha} g_{\beta \rho}
+ \partial_{\beta} g_{\rho \alpha} - \partial_{\rho} g_{\alpha \beta}\right) \, .
\end{eqnarray}
On the other hand, in the Palatini formalism both $g_{\mu\nu}$ and $\Gamma$ are treated as independent variables, and the only constraint is that the connection is torsion-free, $\Gamma^\lambda_{\alpha\beta}=\Gamma^\lambda_{\beta\alpha}$. By solving the equations of motion we obtain \cite{Bauer:2008zj}
\begin{eqnarray}
\Gamma^{\lambda}_{\alpha \beta} = \overline{\Gamma}^{\lambda}_{\alpha \beta}
+ \delta^{\lambda}_{\alpha} \partial_{\beta} \omega(\phi) +
\delta^{\lambda}_{\beta} \partial_{\alpha} \omega(\phi) - g_{\alpha \beta} \partial^{\lambda}  \omega(\phi) \, ,
\label{eq:conn:J}
\end{eqnarray}
where
\begin{eqnarray}
\label{omega}
\omega\left(\phi\right)=\ln\sqrt{f(\phi)} \, .
\end{eqnarray}
Since the connections (\ref{eq:LC}) and (\ref{eq:conn:J}) are different, the metric and Palatini formulations provide indeed two different theories of gravity.
Alternatively we can understand the differences by studying the problem in the Einstein frame via the conformal transformation
\begin{eqnarray}
\label{eq:gE}
g_{\mu \nu}^E = f(\phi) \ g_{\mu \nu} \, .
\end{eqnarray}
In the Einstein frame gravity looks the same in both the formulations (see also eq. (\ref{eq:conn:J})), however the matter sector (in our case $\phi$) behaves differently. Performing the computations \cite{Bauer:2008zj}, the Einstein frame Lagrangian becomes
\begin{equation}
  \sqrt{- g^{E}} \mathcal{L}^{E} = 
  \sqrt{- g^{E}} \Bigg[  - \frac{M_P^2}{2} R  + 
  \frac{(\partial \chi)^{2}}{2} - U(\chi) \Bigg] \, ,
   \label{eq:Einstein:Lagrangian}
\end{equation}
where $\chi$ is canonically normalized scalar field in the Einstein frame, and its scalar potential is given by
\beq
U(\chi) = \frac{V_\text{eff}(\phi(\chi))}{f^{2}(\phi(\chi))} \, .
\label{eq:U:general}
\ee
In the metric case, $\chi$ is derived by integrating the following equation
\begin{equation}
\frac{\partial \chi}{\partial \phi} = \sqrt{\frac{3}{2}\left(\frac{M_P}{f}\frac{\partial f}{\partial \phi}\right)^2+\frac{1}{f}} \, ,  
  \label{eq:dphiE}
\end{equation}
where the first term comes from the transformation of the Jordan frame Ricci scalar and the second from the rescaling of the Jordan frame scalar field kinetic term. 
On the other hand, in the Palatini case, the field redefinition is induced only by the rescaling of the inflaton kinetic term i.e.
\begin{equation}
\frac{\partial \chi}{\partial \phi} = \sqrt{\frac{1}{f}} \, ,  
  \label{eq:dphiP}
\end{equation}
where there is no contribution from the Jordan frame Ricci scalar. Therefore we can see that the difference between the two formalisms in the Einstein frame relies on the different definition of $\chi$ induced by the different non-minimal kinetic term involving $\phi$.

\section{Inflationary results} \label{sec:inflation}

In this section we investigate the phenomenological implications of the non-minimal coupling in eq. (\ref{eq:f:H}). Since a detailed discussion of reheating is beyond the purpose of the present article, we do not need to specify the exact shape of the potential around its minimum. It is sufficient to assume that during inflation the potential is well described by eqs. (\ref{eq:Veff}) and (\ref{eq:l:eff:better}). The corresponding Einstein frame scalar potential is given by
\bea
 U(\chi)  &=& \frac{\lambda_{\rm eff}(\phi,\mu) \, M_P^4 \, \phi(\chi)^4}{4\left[ M_P^2 + \xi \phi(\chi)^2 \right]^2} \nn\\     
 &=& \frac{\lambda(\mu) \, M_P^4 \, \phi(\chi)^4}{4\left[ M_P^2 + \xi \phi(\chi)^2 \right]^2} \left[1 + \frac{9 \lambda (\mu )}{16 \pi ^2} \ln \left(\frac{3 \lambda_0 \phi(\chi)^2}{\mu^2}\right) \right] ,
\label{eq:U}
\eea
where $\lambda(\mu)$ is given in (\ref{eq:l:run}) and the difference between the metric and the Palatini formulations is given by the different solution of eqs. (\ref{eq:dphiE}) and (\ref{eq:dphiP}). 

Assuming slow-roll, the inflationary dynamics is described by the usual slow-roll parameters and the total number of $e$-folds during inflation\footnote{The exact number of $e$-folds is related to the reheating mechanism and it can be used for discriminating between the metric and the Palatini formulations \cite{Racioppi:2017spw}.  Here we concentrate only on the physics during inflation, being the study of reheating beyond the scope of the present article.}. The slow-roll parameters are defined as
\beq
\epsilon \equiv \frac{1}{2}M_{\rm P}^2 \left(\frac{1}{U}\frac{{\rm d}U}{{\rm d}\chi}\right)^2 \,, \quad
\eta \equiv M_{\rm P}^2 \frac{1}{U}\frac{{\rm d}^2U}{{\rm d}\chi^2} \,,
\ee
and the number of $e$-folds as
\beq
N_e = \frac{1}{M_{\rm P}^2} \int_{\chi_f}^{\chi_N} {\rm d}\chi \, U \left(\frac{{\rm d}U}{{\rm d} \chi}\right)^{-1},
\label{Ndef}
\ee
where the field value at the end of inflation, $\chi_f$, is defined via $\epsilon(\chi_f) = 1$.  
The field value $\chi_N$ at the time a given scale left the horizon is given by the corresponding $N_e$. 
To reproduce the correct amplitude for the curvature power spectrum, the potential has to satisfy \cite{Planck2018:inflation}
\beq
\label{cobe}
\ln \left(10^{10} A_s \right) = 3.044 \pm 0.014   \, ,
\ee
where
\beq
 A_s = \frac{1}{24 \pi^2 \MP^4} \frac{U(\chi_N)}{\epsilon(\chi_N)} \label{eq:As}
\ee
and the other two relevant observables, i.e. the spectral index and the tensor-to-scalar ratio are expressed in terms of the slow-roll parameters by
\bea
n_s &\simeq& 1+2\eta-6\epsilon \label{eq:ns} \\
r &\simeq& 16\epsilon ,
\eea
respectively. 
Before performing a detailed numerical analysis, let us discuss the strong coupling limit, $\xi \to +\infty$. In this case the two formulations share a similar Einstein frame field redefinition
\begin{equation}
\chi \simeq \frac{M_P}{q} \left[\frac{8 \pi ^2}{9 \lambda _0}+\ln \left(\frac{\sqrt{3 \lambda _0} \phi }{M_P}\right)\right]
  \label{eq:chi}
\end{equation}
where we set in a convenient way the value $\chi=0$ and $q$ is either
\begin{equation}
q = q_m = \sqrt{\frac{\xi}{1+ 6 \xi}} \, ,
  \label{eq:q:metric}
\end{equation}
for metric gravity, or 
\begin{equation}
q = q_P = \sqrt{\xi} \, ,
  \label{eq:q:Palatini}
\end{equation}
for Palatini gravity. In the strong coupling limit the Einstein frame potential behaves like a running cosmological constant
\bea
 \lim_{\xi \to \infty} U(\chi) &\simeq & \left[ \lim_{\xi \to \infty} \lambda_{\rm eff} (\phi,\mu) \right] \frac{ M_P^4}{4\xi^2} 
\label{eq:U:limit}
\eea
If $\lambda_0$ is small ($\lambda_0 \ll 1$),  we can replace $\lambda_{\rm eff}$ with $\lambda_{\rm app}$ and get
\bea
 \lim_{\xi \to \infty} U(\chi) &\simeq & \frac{\lambda_0 \, M_P^4 \, \phi(\chi)^4}{4\left[ \xi \phi(\chi)^2 \right]^2} \left[1 + \frac{9 \lambda_0}{16 \pi ^2} \ln \left(\frac{3 \lambda_0 \phi(\chi)^2}{\MP^2}\right) \right] \nn\\
 & = & \frac{9 \lambda _0^2}{32 \pi ^2 } \frac{ M_P^3}{\xi ^2} q \, \chi \, .
\label{eq:U:limit:app}
\eea
We can see that in both gravity formulations the limit solution is linear inflation, with the only difference in the normalization factor $q$. As expected, this result is in agreement with \cite{Racioppi:2018zoy} with $\delta = \frac{9 \lambda_0}{8 \pi^2}$. For $\lambda_{\rm eff} = \lambda_{\rm I}$ we have
\bea
 \lim_{\xi \to \infty} \lambda_{\rm I} &=&  \lim_{\xi \to \infty} \left[ \lambda (\mu_{\rm I} ) + 
 \frac{9 \lambda (\mu_{\rm I} )^2}{16 \pi ^2} \ln \left(\frac{3 \lambda_0 \phi ^2 }{\mu_{\rm I}^2}\right) \right]  \simeq 
\frac{\lambda _0}{1-\frac{9 \lambda _0 \ln \left(\frac{3 \lambda _0}{\xi
   }\right)}{16 \pi ^2}}+\frac{9 \lambda _0^2 \ln \left(\frac{\xi  \phi
   ^2}{M_{\text{P}}^2}\right)}{16 \pi ^2 \left[1-\frac{9 \lambda _0 \ln
   \left(\frac{3 \lambda _0}{\xi }\right)}{16 \pi ^2}\right]^2} \nn\\
&=&  \bar\lambda_0 +
 \frac{9 \bar\lambda_0^2}{16 \pi ^2} \ln \left(\frac{3 \bar\lambda_0 \phi^2 }{\bar\mu_0^2}\right)
\label{eq:lambda:I:limit}
\eea
where
\beq
 \bar\lambda_0 =\frac{\lambda _0}{1-\frac{9 \lambda _0 \ln \left(\frac{3 \lambda _0}{\xi
   }\right)}{16 \pi ^2}} \qquad {\rm and} \qquad
 \bar\mu_0 =  \sqrt{\frac{3 \bar{\lambda }_0}{\xi }} M_{\text{P}}
\ee
It is interesting to notice that eq. (\ref{eq:lambda:I:limit}) is the same as $\lambda_{\rm app}$ in eq. (\ref{eq:l:app}) with the replacements $\lambda_0 \to \bar \lambda_0$ and $\mu_0 \to \bar \mu_0$. Therefore we expect that inflationary results for $\lambda_{\rm eff}=\lambda_{\rm app,I}$ will be the same in the strong coupling limit, but for different values of $\lambda_0$ and $\xi$. And again this holds independently on the formulation of gravity. Finally, for $\lambda_{\rm eff}=\lambda_{\rm II}$ we have
\bea
 \lim_{\xi \to \infty} \lambda_{\rm II} &=&  \lambda (\mu_{\rm II} ) + 
 \frac{9 \lambda (\mu_{\rm II} )^2}{16 \pi ^2} \ln \left(\frac{3 \lambda_0 \phi ^2 }{\mu_{\rm II}^2}\right)   = \frac{\lambda _0}{1-\frac{9 \lambda _0 \ln \left(\frac{3 \lambda _0 \phi
   ^2}{M_{\text{P}}^2}\right)}{16 \pi ^2}}
\label{eq:lambda:II:limit}
\eea
and therefore
\bea
 \lim_{\xi \to \infty} U(\chi) &\simeq & \frac{\lambda _0}{1-\frac{ 9 \lambda _0 }{8 \pi ^2} \frac{q \, \bar\chi }{ \MP}} \frac{ M_P^4}{4\xi^2} \, ,
\label{eq:U:II:limit}
\eea
where we used eq. (\ref{eq:chi}) and
\beq
\bar\chi = \chi - \frac{M_P}{q} \frac{8 \pi ^2}{9 \lambda _0} \, .
\ee
In this case the potential does not resemble the behaviour of linear inflation, in contrast to what happens with $\lambda_{\rm eff}=\lambda_{\rm app,I}$. Moreover by using eq. (\ref{eq:ns}) we get
\beq
 n_s \simeq 1+ \left[ \frac{9 \, q \, \lambda _0}{8 \pi ^2 \left(1-\frac{9 \lambda _0 q \bar\chi }{8
   \pi ^2 M_{\text{P}}}\right)} \right]^2\, , \label{eq:ns:limit:II}
\ee
which means that $n_s \gtrsim 1$ and the strong coupling limit of $\lambda_{\rm eff}=\lambda_{\rm II}$ is ruled out.

For completeness, we perform a full inflationary analysis considering also $\xi$ values not in the strong coupling limit. We proceed in the following way. We first fix the gravity formulation (metric or Palatini) and then the effective coupling that we want to study ($\lambda_{\rm eff}=\lambda_{\rm app},\lambda_{\rm I},\lambda_{\rm II}$). Assuming $N_e=50$, we remain with only two free parameters: $\lambda_0$ and $\xi$. We vary $\lambda_0$ between $\lambda_0 \approx 10^{-13}$ (the usual value for quartic inflation) and $\lambda_0=1$ (naive upper limit set as a necessary, but not always sufficient, condition to ensure perturbativity of the theory during inflation). Therefore $\xi$ is fixed in order to satisfy the constraint (\ref{cobe}). 

\begin{figure}[p!]
\begin{center}
 \includegraphics[width=0.49\textwidth]{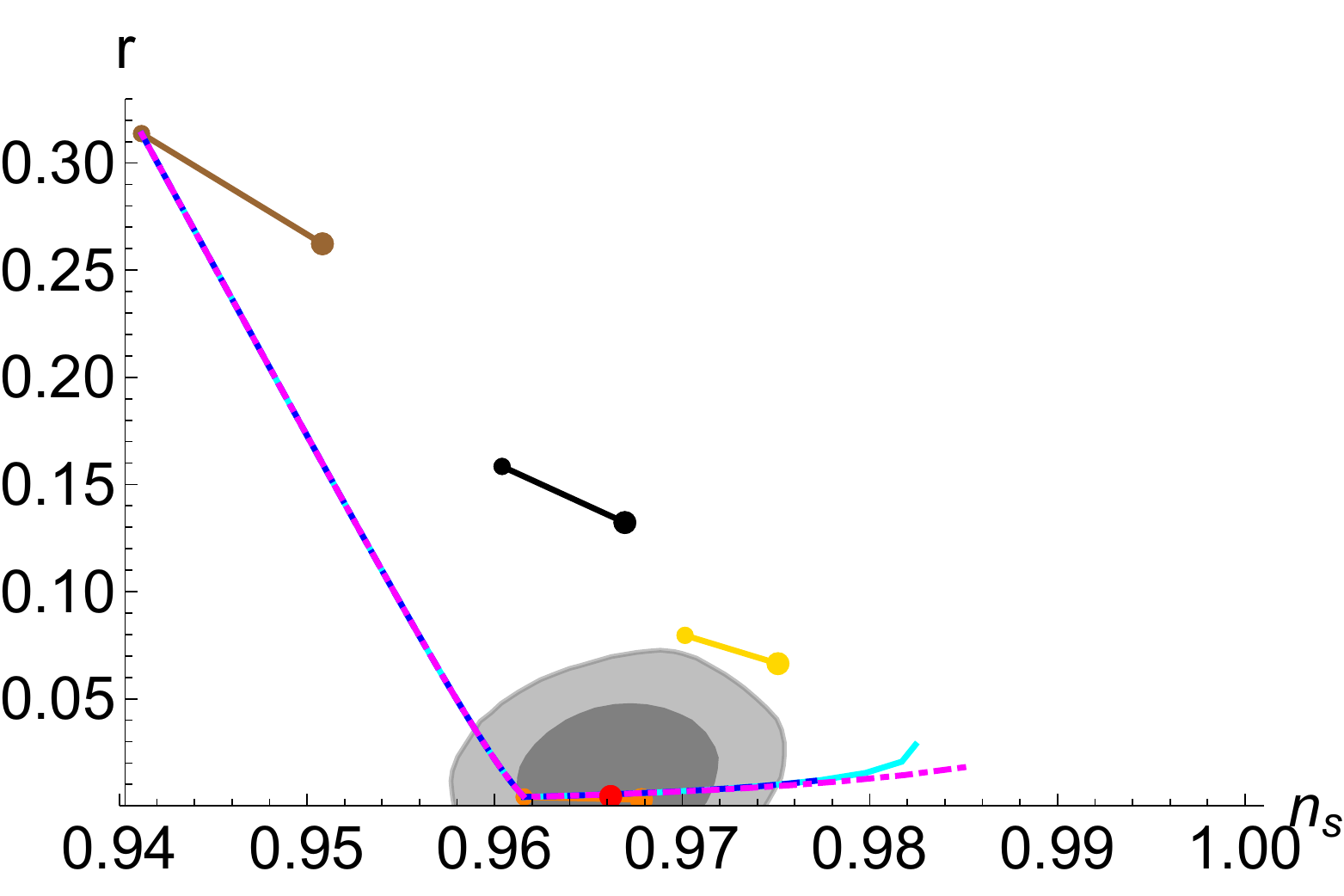}
 \includegraphics[width=0.49\textwidth]{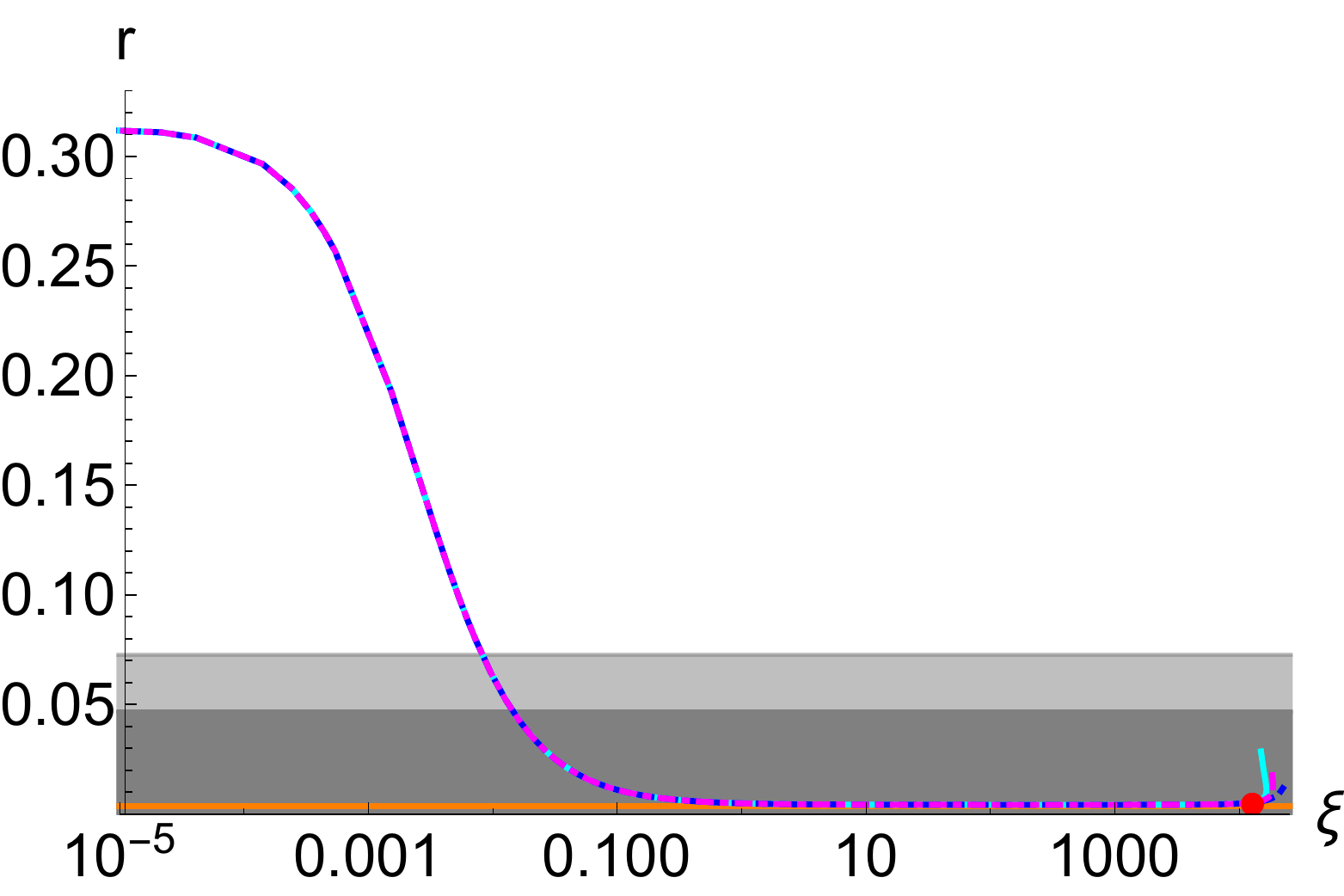}\\
 a) \hspace{0.45\textwidth} b) \\
 \includegraphics[width=0.49\textwidth]{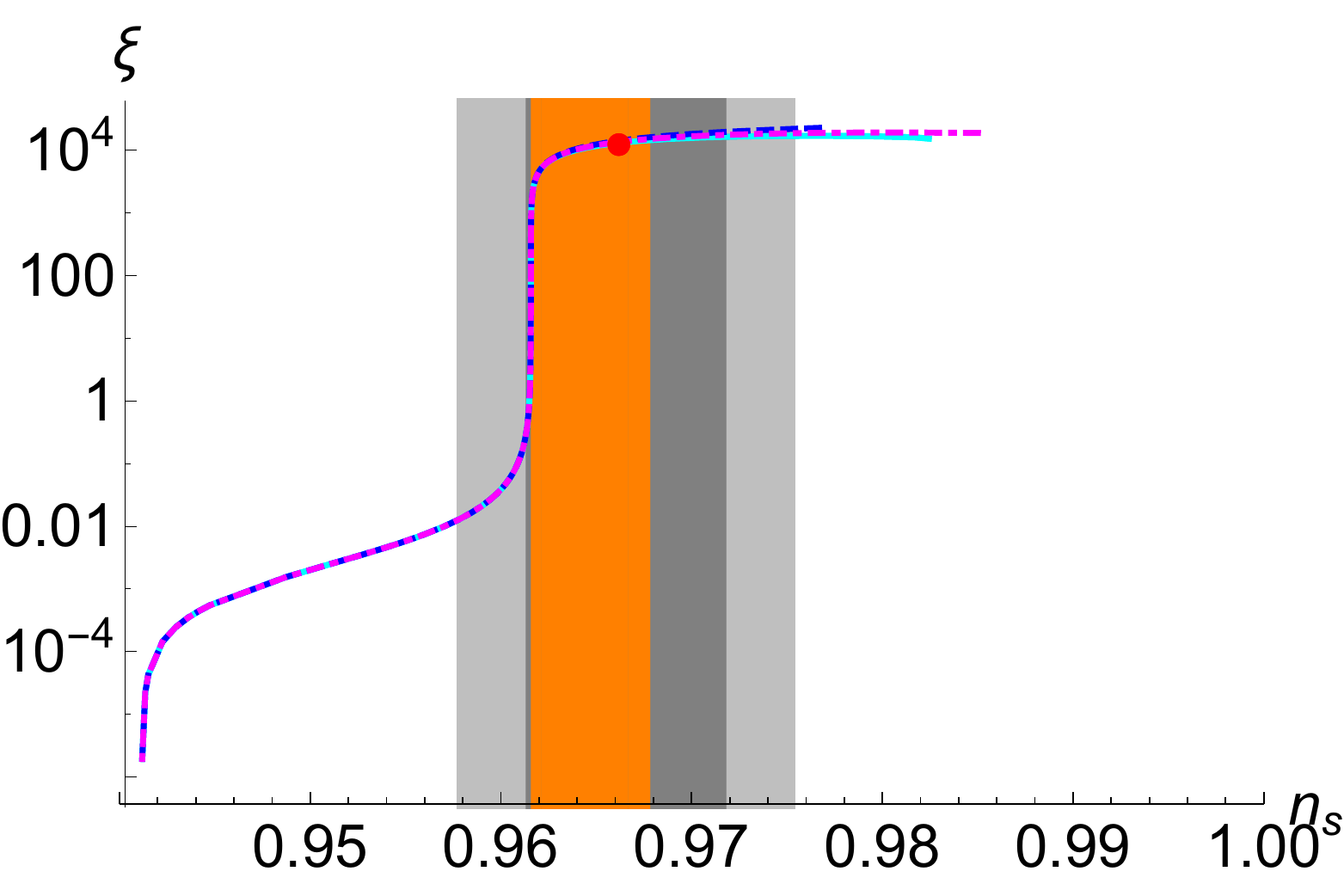}
 \includegraphics[width=0.49\textwidth]{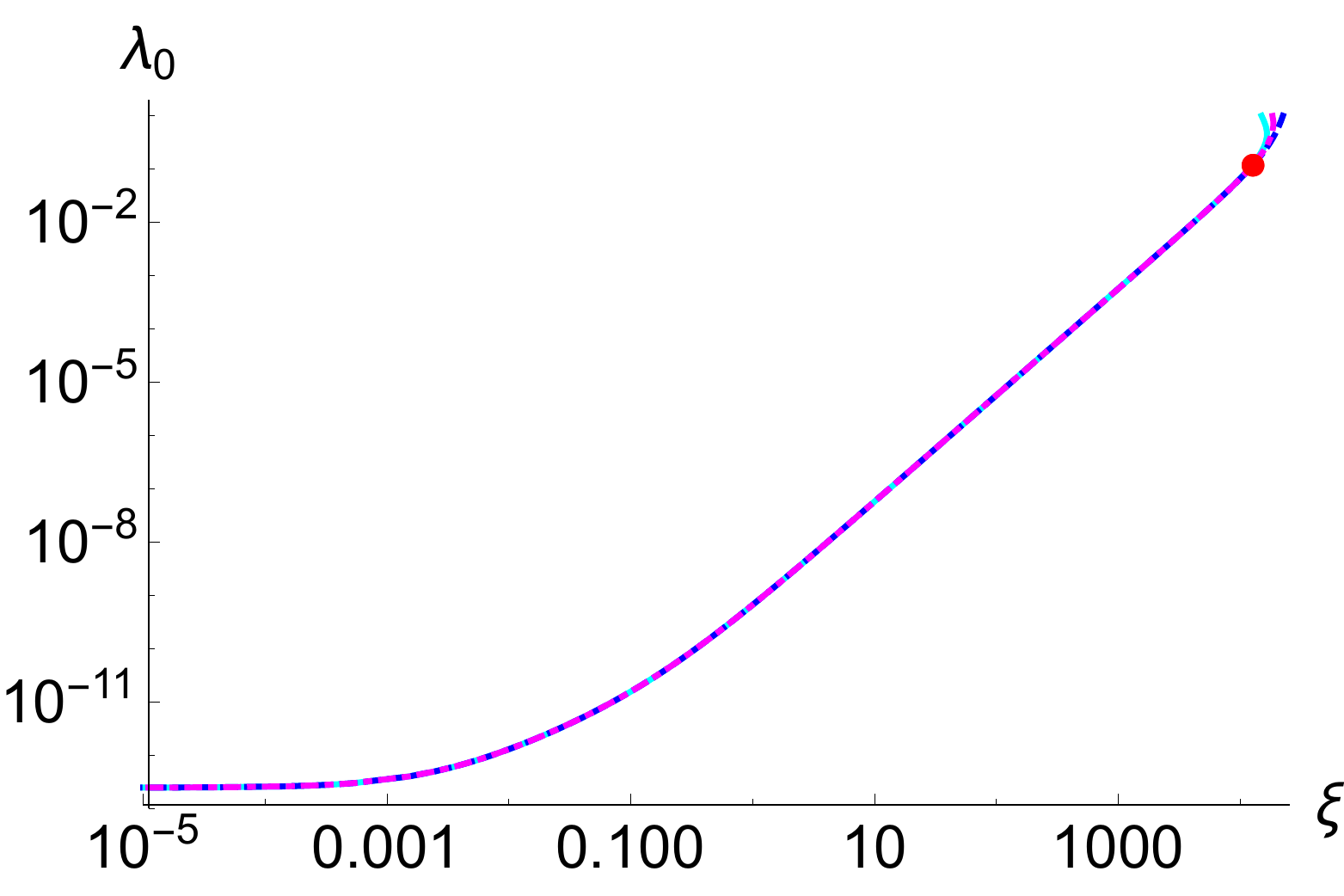}\\
 c) \hspace{0.45\textwidth} d)
\end{center}
 \caption{Metric formulation: $r$ vs. $n_s$ (a), $r$ vs. $\xi$ (b), $\xi$ vs. $n_s$ (c) and $\lambda_0$ vs. $\xi$ (d) for $\lambda_{\rm eff}=\lambda_{\rm app}$ (cyan), $\lambda_{\rm eff}=\lambda_{\rm I}$ (blue, dashed) and $\lambda_{\rm eff}=\lambda_{\rm II}$ (magenta, dot-dashed) with $N_e =50$ $e$-folds. The red point indicates the loss of accuracy of the present analysis. For reference we also plot predictions of quartic (brown), quadratic (black) and linear (yellow) inflation for $N_e \in [50,60]$. The gray areas represent the 1,2$\sigma$ allowed regions coming from Planck 2018 data~\cite{Planck2018:inflation}.
 }
  \label{Fig:Results:metric}
\end{figure}

\begin{figure}[p!]
\begin{center}
 \includegraphics[width=0.49\textwidth]{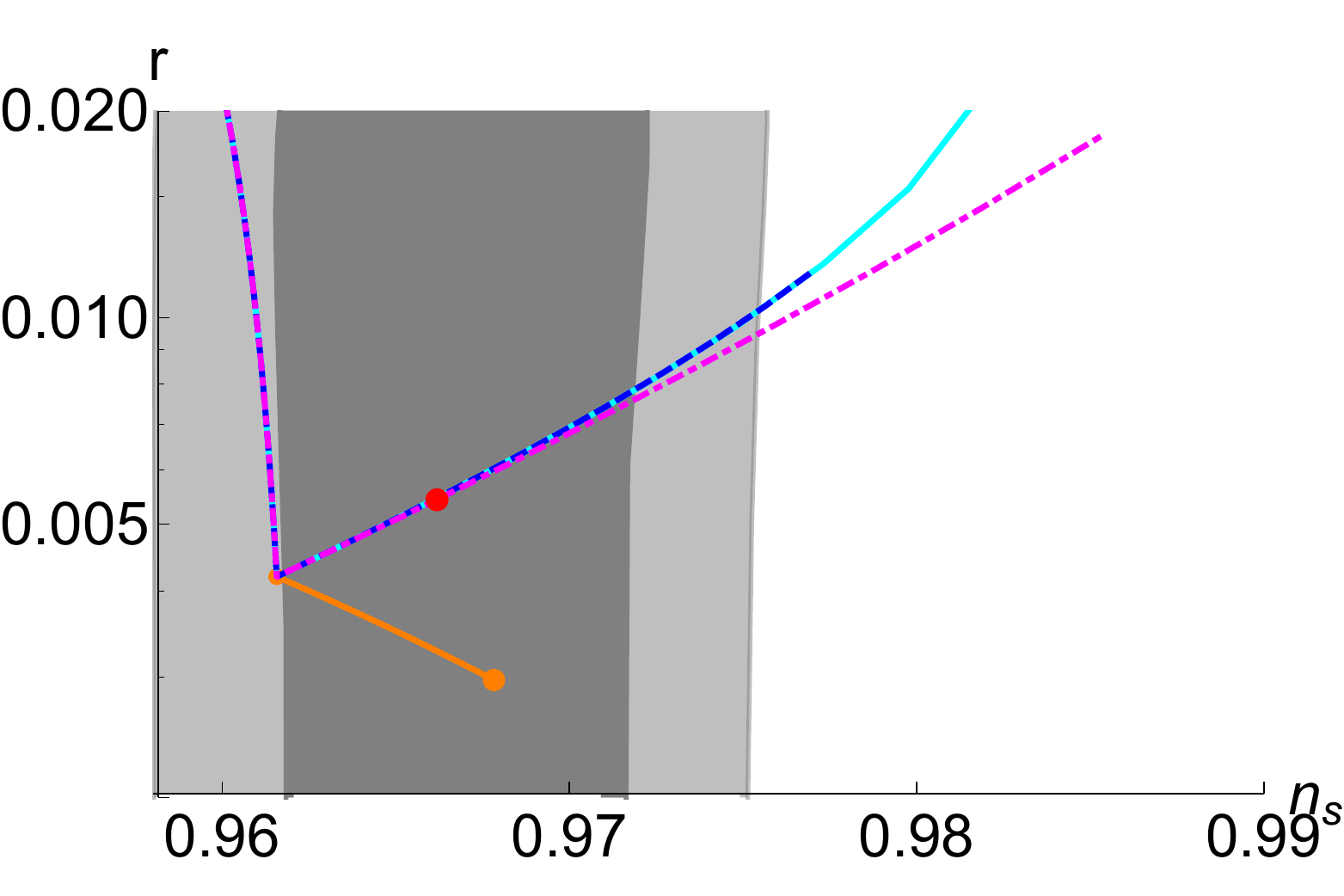}
 \includegraphics[width=0.49\textwidth]{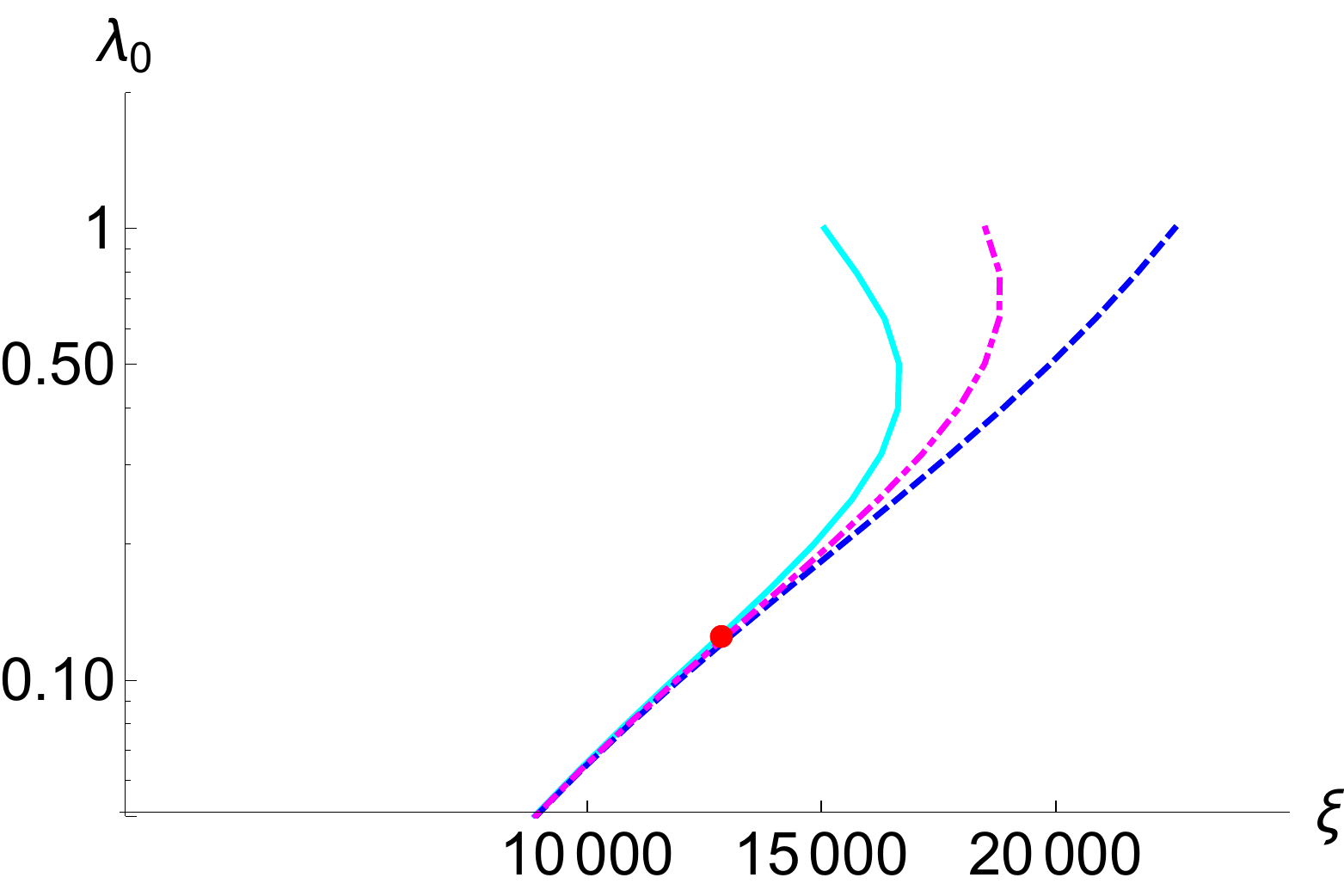}\\
 a) \hspace{0.45\textwidth} b)
\end{center}
 \caption{Metric formulation: zoom of $r$ vs. $n_s$ (a) for $r\leq 0.02$ and $\lambda$ vs. $\xi$ (b) for $\lambda_0 \geq 0.05$. The color code is the same as in Fig. \ref{Fig:Results:metric}. }
  \label{Fig:Results:metric:zoom}
\end{figure}

\begin{figure}[p!]
\begin{center}
 \includegraphics[width=0.49\textwidth]{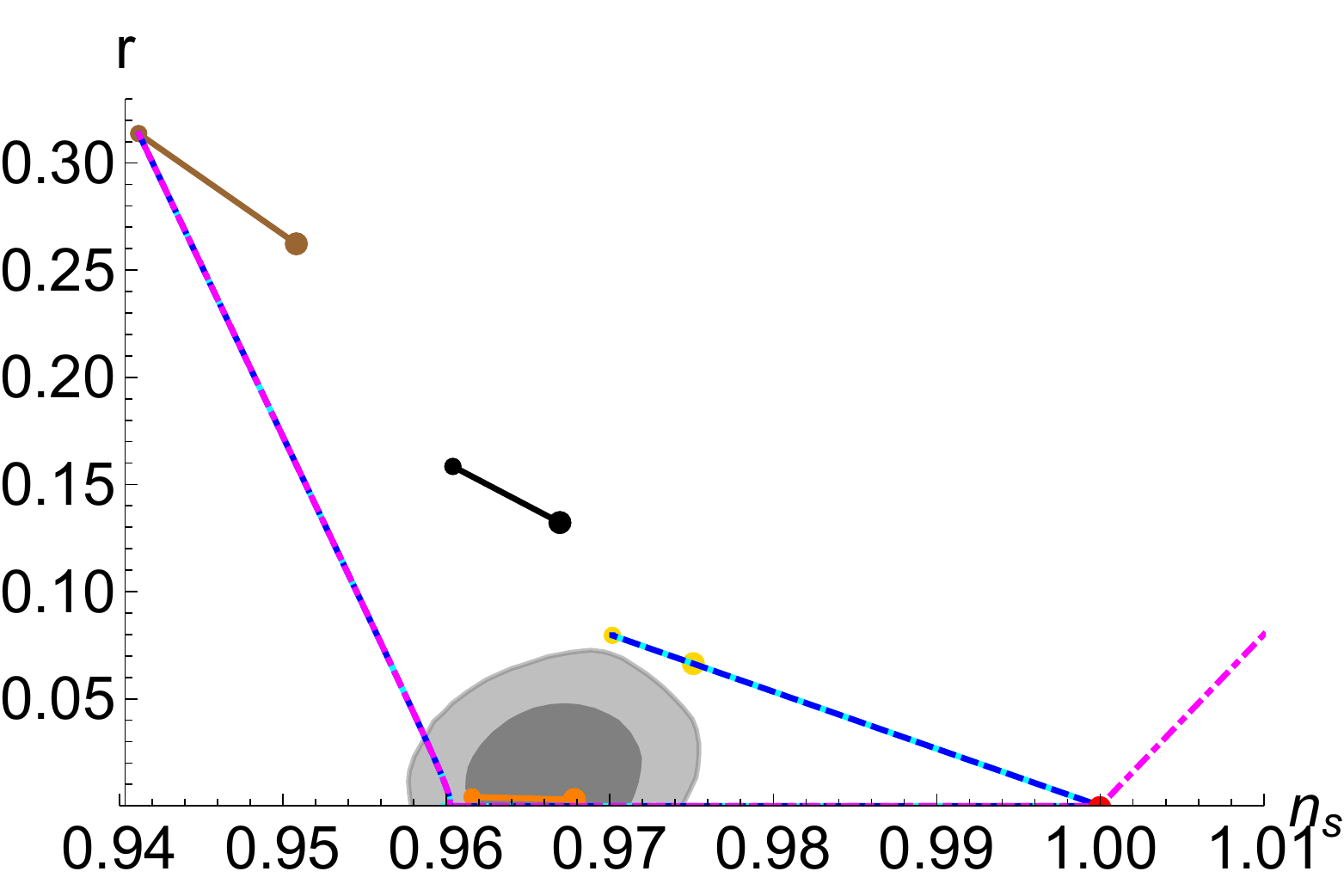}
 \includegraphics[width=0.49\textwidth]{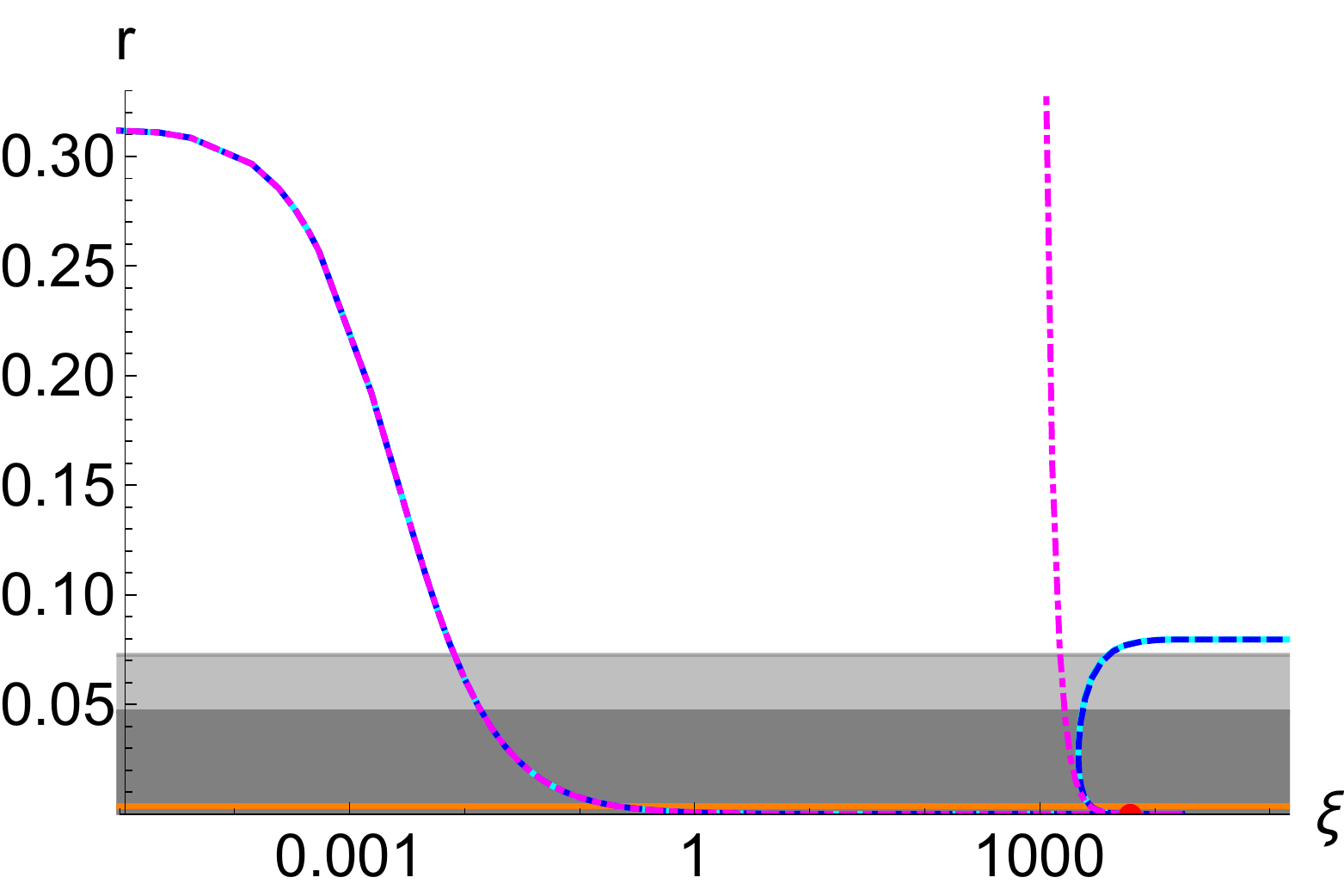}\\
 a) \hspace{0.45\textwidth} b)\\
 \includegraphics[width=0.49\textwidth]{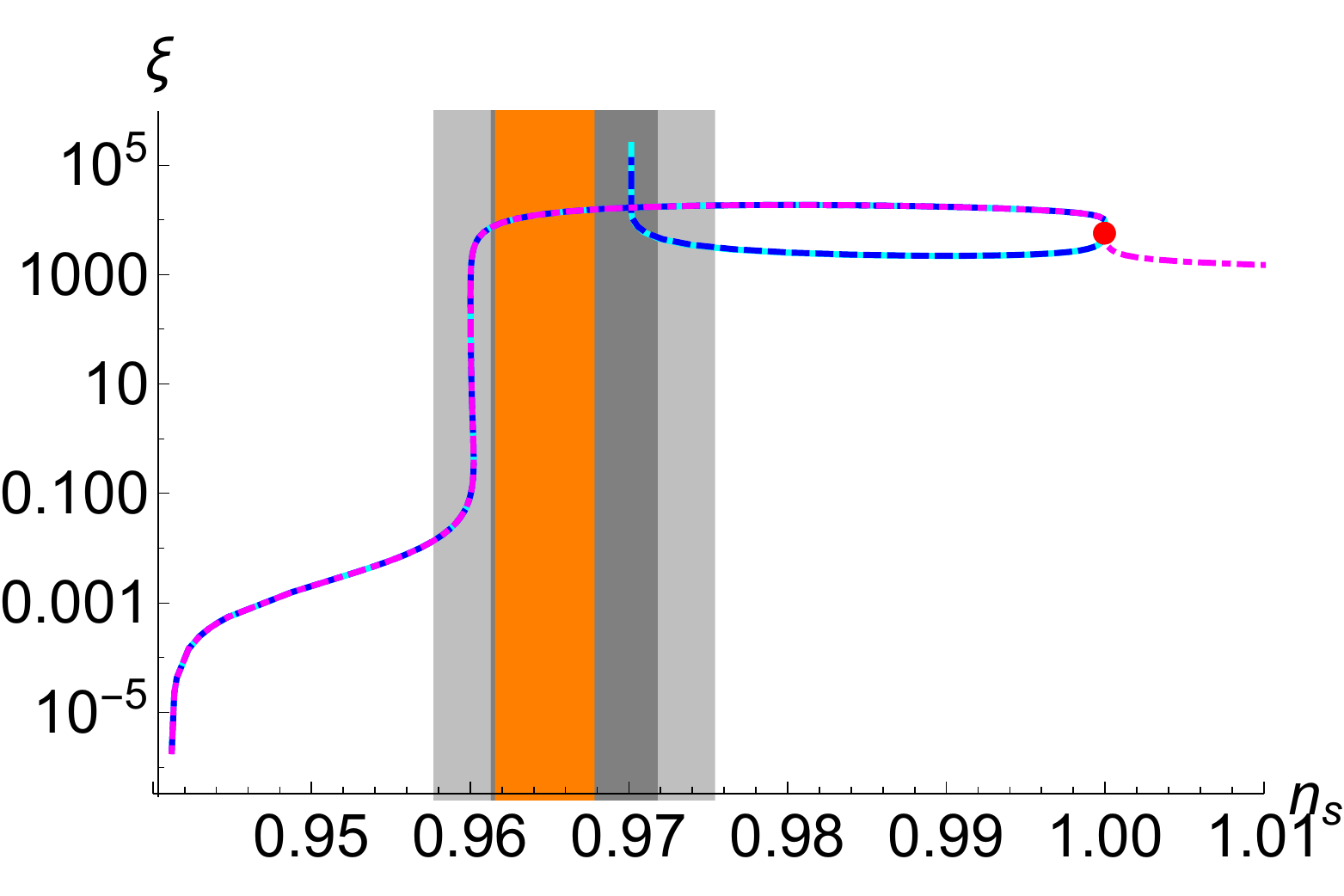}
 \includegraphics[width=0.49\textwidth]{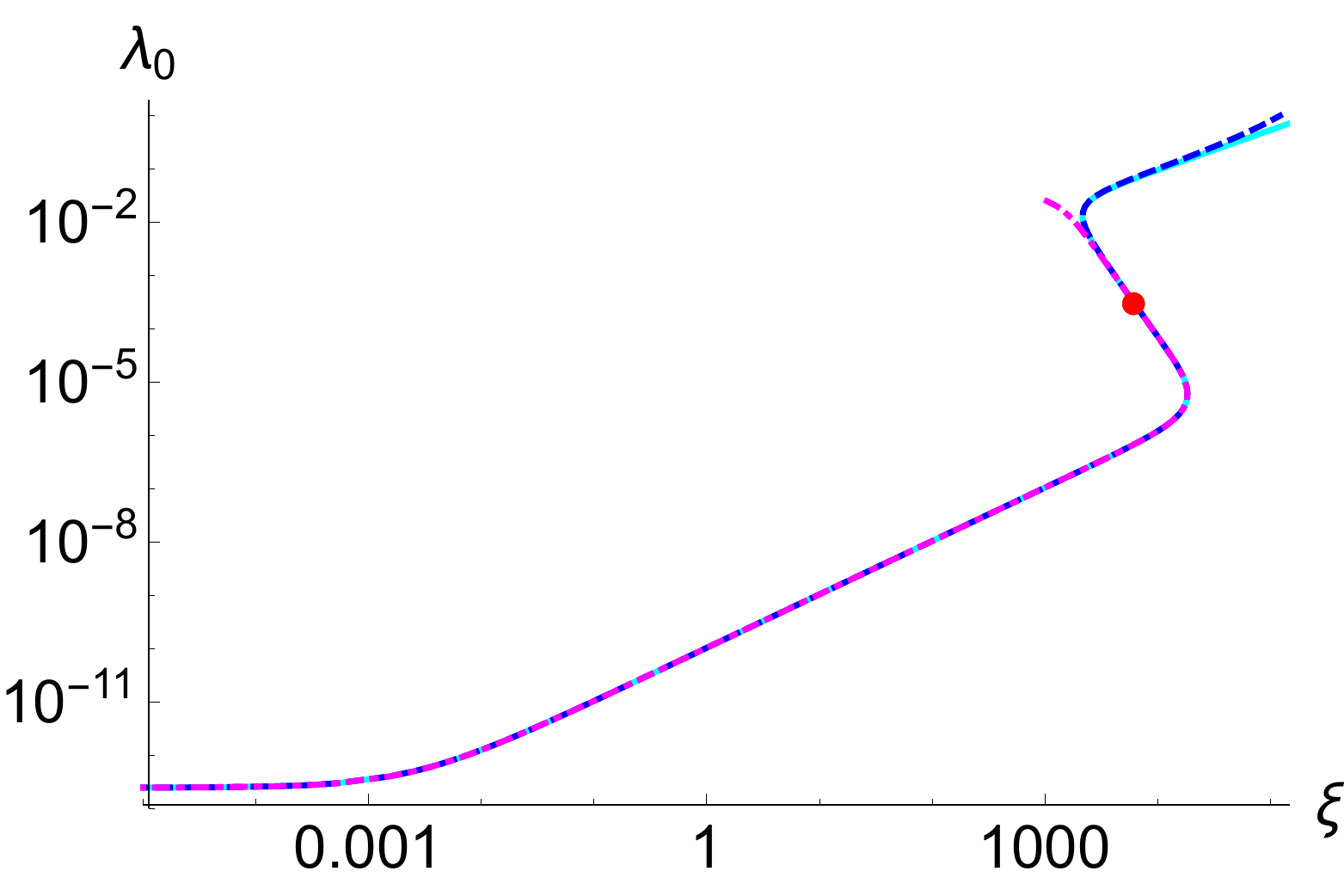}\\
 c) \hspace{0.45\textwidth} d)
\end{center}
 \caption{Palatini formulation: $r$ vs. $n_s$ (a), $r$ vs. $\xi$ (b), $\xi$ vs. $n_s$ (c) and $\lambda_0$ vs. $\xi$ (d) for $\lambda_{\rm eff}=\lambda_{\rm app}$ (cyan), $\lambda_{\rm eff}=\lambda_{\rm I}$ (blue, dashed) and $\lambda_{\rm eff}=\lambda_{\rm II}$ (magenta, dot-dashed)  with $N_e =50$ $e$-folds.  The red point indicates the loss of accuracy of the present analysis. For reference we also plot predictions of quartic (brown), quadratic (black) and linear (yellow) inflation for $N_e \in [50,60]$. The gray areas represent the 1,2$\sigma$ allowed regions coming from Planck 2018 data~\cite{Planck2018:inflation}.}
  \label{Fig:Results:Palatini}
\end{figure}

\begin{figure}[p!]
\begin{center}
 \includegraphics[width=0.49\textwidth]{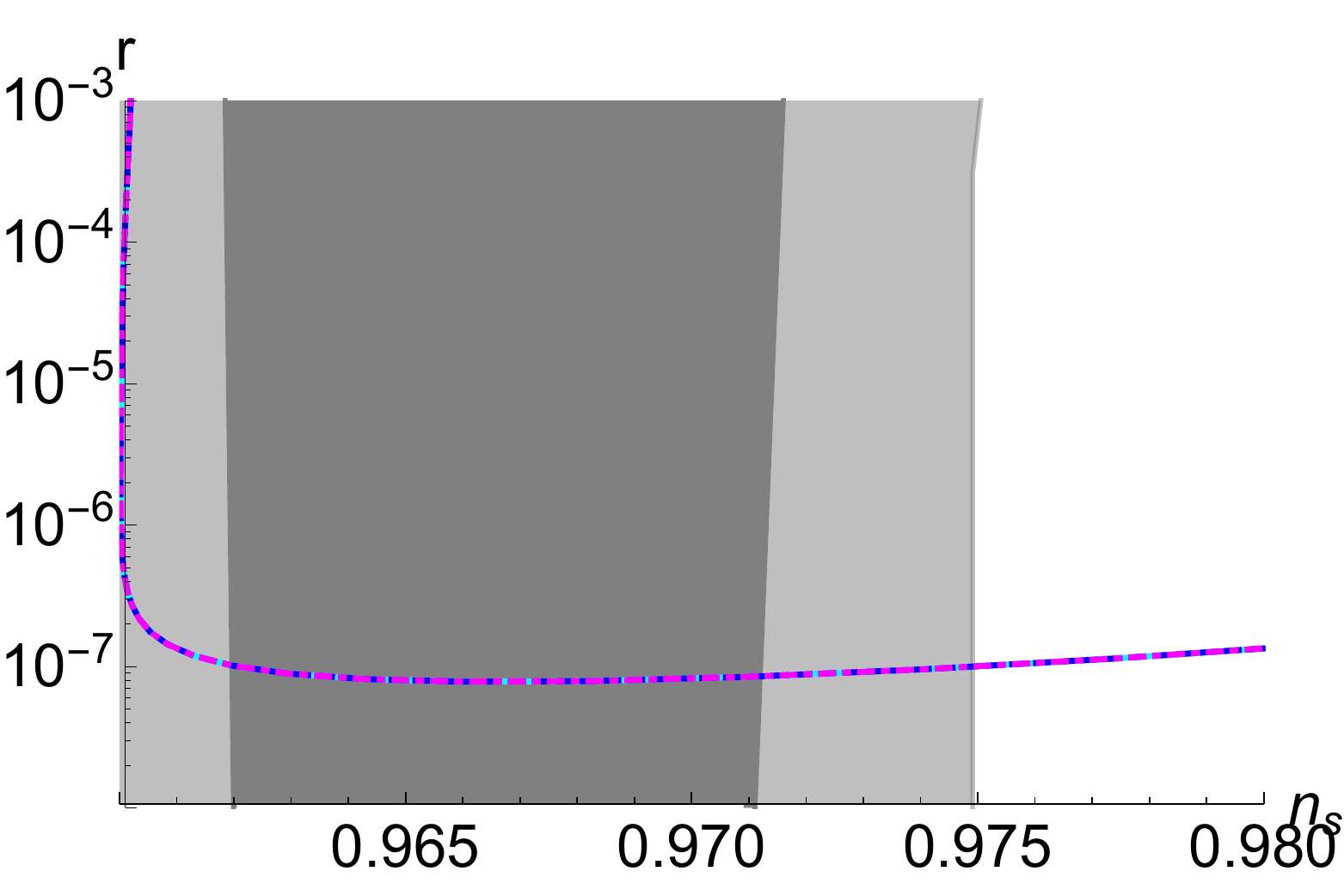}
 \includegraphics[width=0.49\textwidth]{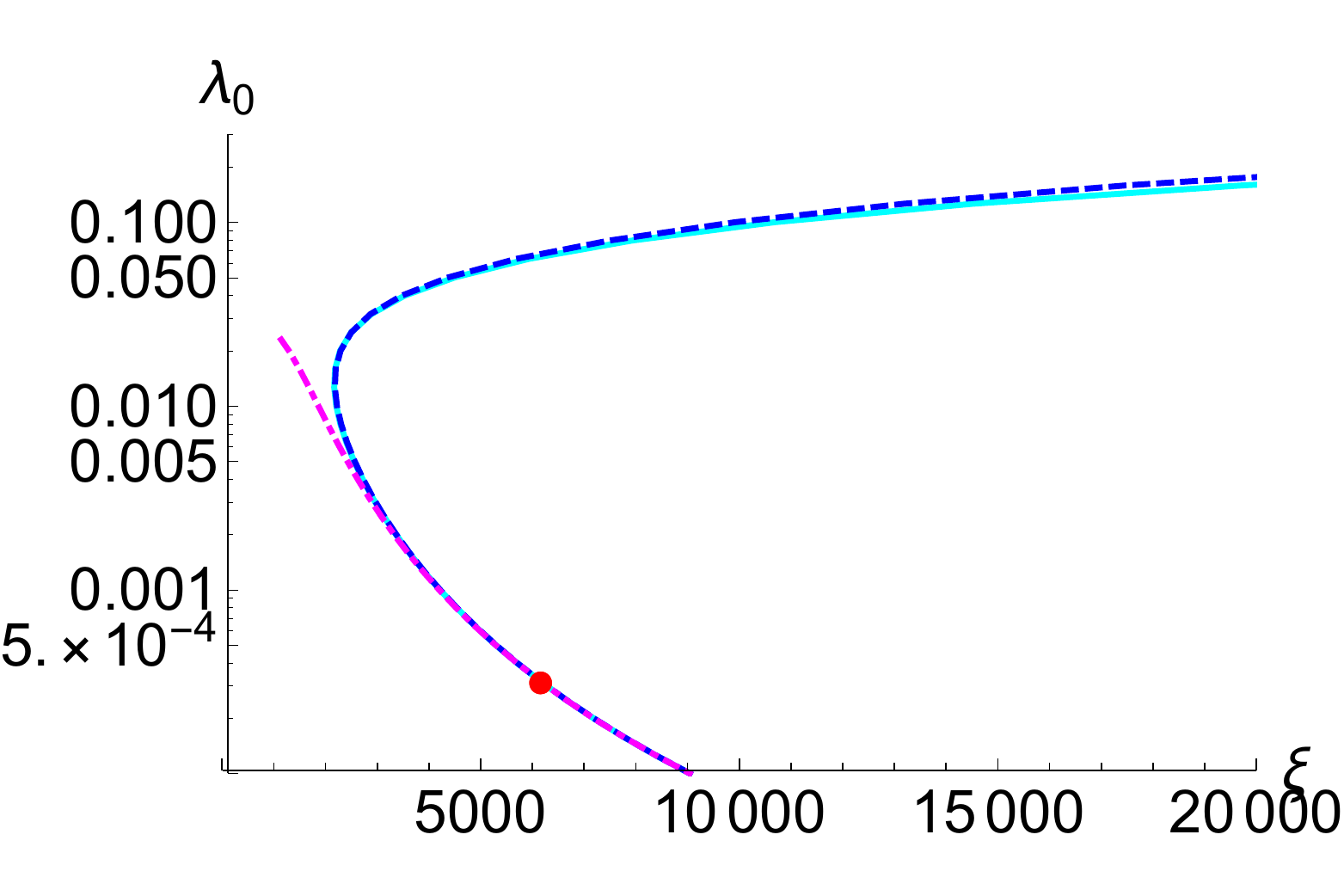}\\
 a) \hspace{0.45\textwidth} b)
\end{center}
 \caption{Palatini formulation: zoom of $r$ vs. $n_s$ (a) for $r\leq 10^{-3}$  and $\lambda_0$ vs. $\xi$ (b) for $\lambda_0 \leq 0.3$. The color code is the same as in Fig. \ref{Fig:Results:Palatini}. }
  \label{Fig:Results:Palatini:zoom}
\end{figure}

The corresponding results are given in Figs.~\ref{Fig:Results:metric}, \ref{Fig:Results:metric:zoom}, \ref{Fig:Results:Palatini} and \ref{Fig:Results:Palatini:zoom}.
In Fig.~\ref{Fig:Results:metric} we are presenting the results for the metric formulation and plotting $r$ vs. $n_s$ (a), $r$ vs. $\xi$ (b), $\xi$ vs. $n_s$ (c) and $\lambda_0$ vs. $\xi$ (d) for $\lambda_{\rm eff}=\lambda_{\rm app}$ (cyan), $\lambda_{\rm eff}=\lambda_{\rm I}$ (blue, dashed) and $\lambda_{\rm eff}=\lambda_{\rm II}$ (magenta, dot-dashed)  with $N_e =50$ $e$-folds. For reference we also plot predictions of quartic (brown), quadratic (black) and linear (yellow) inflation for $N_e \in [50,60]$. The gray areas represent the 1,2$\sigma$ allowed regions coming from Planck 2018 data~\cite{Planck2018:inflation}. In Fig. \ref{Fig:Results:metric:zoom} we show the same plots for $r$ vs. $n_s$ (a) and $\lambda_0$ vs. $\xi$ (b) as in Fig. \ref{Fig:Results:metric}, but respectively zoomed in the regions $r\leq 0.02$ and $\lambda_0 \geq 0.05$. Figs. \ref{Fig:Results:Palatini} and \ref{Fig:Results:Palatini:zoom} are the same as Figs.~\ref{Fig:Results:metric} and \ref{Fig:Results:metric:zoom} but for the Palatini formulation of gravity and a zoom respectively for $r\leq 10^{-3}$ and $\lambda_0 \leq 0.3$.

The results of the two formulations share some similarities.  First, for $\xi \simeq 0$, the predictions are compatible with the ones of standard quartic inflation. Then, by increasing $\xi$ until $\xi \lesssim 10^3$, the predictions are aligned with the respective strong-coupling limits of the standard (without loop corrections) non-minimal inflation (\cite{Kallosh:2013tua} for metric and \cite{Jarv:2017azx} for Palatini). The attractor limit is well described in the vertical region around $n_s \simeq 0.96$ in Figs. \ref{Fig:Results:metric}c and \ref{Fig:Results:Palatini}c.
When $\lambda_0$ (or equivalently $\xi$) is small enough, the predictions for  $\lambda_{\rm eff}=\lambda_{\rm app,I,II}$ are overlapping. This happens for $\lambda_0 \lesssim 0.1$ $ (\xi \lesssim 1.2 \times 10^4)$ in metric gravity and $\lambda_0 \lesssim 10^{-3}$ $(\xi \lesssim 4 \times 10^3)$ in Palatini gravity.
Moreover, as anticipated before, it is impossible to discriminate between $\lambda_{\rm eff}=\lambda_{\rm app},\lambda_{\rm I}$ in the $r$ vs. $n_s$ plots respectively in both gravity formulations for any values of $\lambda_0$ and $\xi$. Differences are appreciable only in the actual values of those two parameters. This happens for $\lambda_0 \gtrsim 0.2$ in metric gravity and $\lambda_0 \gtrsim 0.01$ in Palatini gravity.

On the other hand, there are several differences between the results of the two formulations. First of all, while in the Palatini formulation it is possible to reach the linear inflation limit \cite{Racioppi:2018zoy} within perturbativity, this never happens in case of metric gravity.  The same happens for the strong coupling limit of $\lambda_{\rm eff}=\lambda_{\rm II}$, which is only allowed in Palatini gravity.
Moreover, the lower limit for $r$ in the metric case coincides with the prediction of $R^2$ inflation, where quantum corrections are still sub-dominant, while for the Palatini case the lower limit is $r \gtrsim 10^{-7}$, where the loop effects are relevant. 
Furthermore, only  in the Palatini formulation for $\lambda_{\rm eff}=\lambda_{\rm II}$ with $\lambda_0 \gtrsim 0.03 $ we encountered Landau poles in the inflationary region and therefore removed the corresponding points.
In addition, comparing Figs. \ref{Fig:Results:metric} and \ref{Fig:Results:Palatini} we can see that while in the metric case $\xi$ is monotonically increasing with $\lambda_0$, in the Palatini one $\xi$ increases until $n_s \sim 1$ is reached, then it decreases, and then it increases again. In the first and third region $\xi$ behaves as expected, therefore let us focus on the region $n_s \lesssim 1$, where there results of $\lambda_0$ vs. $\xi$ were unforeseen. As shown in Fig. \ref{Fig:Results:Palatini}d, in this region the lines of $\lambda_{\rm eff}=\lambda_{\rm app,I,II}$ are still overlapped, therefore it is enough to study the $\lambda_{\rm eff}=\lambda_{\rm app}$ case. In such a region we can still apply the strong coupling limit $\xi \to \infty$ and  approximate the Einstein frame potential again as a running cosmological constant (see eq. (\ref{eq:U:limit})). However now it is convenient to solve the field redefinition (\ref{eq:dphiP}) as follows
\beq
\chi \simeq \frac{M_{\text{P}} }{\sqrt{\xi }} \left[\frac{1}{2}\ln \left(\frac{3 \lambda _0}{4 \xi }\right)+\ln    \left(\frac{2\sqrt{\xi } \phi }{M_{\text{P}}}\right)\right]
\ee
where we conveniently shifted the position of $\chi=0$. Therefore the Einstein frame potential becomes
\beq
 U(\chi) \simeq  \frac{\lambda _0 M_{\text{P}}^4 }{4 \xi ^2} \left(1+\frac{9 \lambda _0 \sqrt{\xi } \chi }{8   \pi ^2 M_{\text{P}}}\right) \, .
\ee
It can proven that in this case $\frac{9 \lambda _0 \sqrt{\xi } \chi }{8   \pi ^2 M_{\text{P}}} \ll 1$ and therefore we can get the following approximated results
\begin{equation}
 r  \simeq r_0 \left( 1-\frac{r_0}{4} N_e  \right)  \, , \quad
 n_s  \simeq  1-\frac{3}{8} r \, , \label{eq:r:and:ns:approx}
\end{equation}
where $r_0$ is the tensor-to-scalar ratio computed at the lowest order
\begin{equation}
 r_0 = \frac{81  \xi }{8 \pi ^4} \lambda _0^2\, .
\end{equation}
The constraint on the amplitude of the perturbation (\ref{cobe}) implies
\beq
A_s \simeq \frac{4 \pi ^2}{243 \lambda _0 \xi ^3} + \frac{\lambda _0 N_e}{16 \pi ^2 \xi ^2} \, .
\ee
Now considering the small $\lambda_0$ limit, we get
\beq
A_s \simeq \frac{4 \pi ^2}{243 \lambda _0 \xi ^3} \, , \label{eq:As:0order}
\ee
and consequently
\begin{equation}
r_0 \simeq \frac{2}{729} \frac{1}{A_s^2 \, \xi^5 } \label{eq:r0:As}
\, .
\end{equation}
Henceforth, inserting eq. (\ref{eq:r0:As}) into eq. \eqref{eq:r:and:ns:approx}, we find that, for big enough $\xi$, $r \simeq 0$ and $n_s \simeq 1$.
Moreover, from eq. (\ref{eq:As:0order}) we can see that $\lambda_0$ is inversely proportional to $\xi^3$. 
These results are in agreement with our numerical ones for $\lambda_{\rm eff}=\lambda_{\rm app}$ when $n_s \simeq 1$ in the Palatini formulation. On the other hand it is also interesting to see separately the corresponding limit of the Einstein frame potential for $\lambda_{\rm eff}=\lambda_{\rm II}$. Such limit is already given in eq. (\ref{eq:U:II:limit}) and the consequent constraint on the amplitude is
\beq
A_s \simeq \frac{4 \pi ^2}{243 \lambda _0 \xi ^3}-\frac{\lambda _0 N_e}{48 \pi ^2 \xi ^2} \, , 
\ee
which recovers the results of $\lambda_{\rm eff}=\lambda_{\rm app}$ for small $\lambda_0$ and departs from them by increasing $\lambda_0$ and allowing the possibility of $n_s \gtrsim 1$ (see eq. (\ref{eq:ns:limit:II})). Therefore, in the Palatini formulation it is possible to discriminate between $\lambda_{\rm eff}=\lambda_{\rm I,II}$ far away from the 2$\sigma$ allowed region, nearby $n_s \sim 1$ ($\lambda_0 \gtrsim 0.005$) (represented by the red point in Figs. \ref{Fig:Results:Palatini} and \ref{Fig:Results:Palatini:zoom}). On the other hand, in the metric case it is impossible to discriminate between $\lambda_{\rm eff}=\lambda_{\rm I,II}$ within the 1$\sigma$ region, but it is possible within the 2$\sigma$ boundary from $\lambda_0 \gtrsim 0.15$ $(\xi \gtrsim 1.4 \times 10^4)$  (represented by the red point in Figs. \ref{Fig:Results:metric} and \ref{Fig:Results:metric:zoom}). As we mentioned before, there should be no physical difference in  $\lambda_{\rm eff}=\lambda_{\rm I,II}$, therefore this should be interpreted as a loss of accuracy in the expansion for the effective potential in eq. (\ref{eq:Veff}) and the need to consider higher order loop corrections. 

Finally we conclude remarking that, in agreement with the findings of \cite{Jinno:2019und}, the impact of radiative corrections is stronger in the Palatini formulation rather than in the metric formulation, because the Jordan frame field excursion is larger in the Palatini formulation.

\section{Conclusions} \label{sec:Summary}
Even though we might expect the presence of multiple scalar fields at high energies which will affect the phenomenology, for instance, by inducing multi-field inflationary scenarios (e.g. \cite{Kaiser:2013sna,Kallosh:2013daa,Kuusk:2015dda,Carrilho:2018ffi} and refs therein) and/or relevant non-gaussianities (e.g. \cite{Seery:2005gb} and refs therein), single field models of inflation are attractive for their simplicity and predictivity. 

In particular, in this article we studied a model of quartic inflation where the inflaton field $\phi$ is subject to relevant self-induced radiative corrections and it is coupled non-minimally to gravity. We considered the Higgs-inflation-like non-minimal coupling. We studied the predictions of two different formulations of gravity, metric or Palatini, and the three possible versions of the effective quartic couplings $\lambda_{\rm eff}(\phi,\mu)=\lambda_{\rm app,I,II} (\phi)$: $\lambda_{\rm app}$ is the case in which the tree-level quartic coupling $\lambda_0$ is very small,  we can Taylor expand and explicitly remove from $\lambda_{\rm eff}$ the dependence on the renormalization scale $\mu$, while $\lambda_{\rm eff}(\phi,\mu)=\lambda_{\rm I,II} (\phi)$ corresponds to the prescription I,II choices given in eqs. (\ref{eq:PI}) and (\ref{eq:PII}).
We showed that the formulations share several differences, as expected, but also some interesting similarities. We start with the last ones.

First of all, trivially, the predictions are compatible with the ones of standard quartic inflation  for $\xi \simeq 0$. 
Then, by increasing $\xi$ until $\xi \lesssim 10^3$, the predictions are substantially the same as the respective strong-coupling limits of the standard (tree-level) non-minimal inflation (\cite{Kallosh:2013tua} for metric and \cite{Jarv:2017azx} for Palatini). 
When $\lambda_0 \lesssim 0.1$ $ (\xi \lesssim 1.2 \times 10^4)$ in metric gravity, the predictions for  $\lambda_{\rm eff}=\lambda_{\rm app,I,II}$ are undistinguishable.  The same holds for $\lambda_0 \lesssim 10^{-3}$ $(\xi \lesssim 4 \times 10^3)$ in Palatini gravity.
Moreover, we showed that for $\xi \gg 1$, $\lambda_{\rm app}$ and $\lambda_{\rm I}$ can be mapped into each other just by varying $\lambda_0$ and $\xi$. Therefore it is impossible to distinguish between $\lambda_{\rm eff}=\lambda_{\rm app,I}$ in the $r$ vs. $n_s$ plots respectively in both gravity formulations. Eventual differences are appreciable only in the actual values of $\lambda_0$ and $\xi$.  

On the other hand, the first difference that we notice is the possibility to reach within perturbativity the linear inflation limit \cite{Racioppi:2018zoy} or the strong coupling limit (\ref{eq:U:II:limit}) of $\lambda_{\rm II}$ only in Palatini gravity. Moreover, the lower limit for $r$ in the metric case coincides with the prediction of Starobinsky inflation, while for the Palatini case the lower limit is $r \gtrsim 10^{-7}$. In the metric case $\xi$ is everywhere monotonically increasing with $\lambda_0$, while in the Palatini one $\xi$ can also decrease  around the value $n_s \simeq 1$. Around such a value, still in Palatini gravity, it is also possible to discriminate between $\lambda_{\rm eff}=\lambda_{\rm I,II}$ for $\lambda_0 \gtrsim 0.005$. On the other hand, in the metric case it is impossible to discriminate between $\lambda_{\rm eff}=\lambda_{\rm I,II}$ within the 1$\sigma$ region, but it is possible within the 2$\sigma$ boundary from $\lambda_0 \gtrsim 0.15$ $(\xi \gtrsim 1.4 \times 10^4)$. As there should be no physical difference in  $\lambda_{\rm eff}=\lambda_{\rm I,II}$, this should point out a loss of accuracy in the expansion for the effective potential in eq. (\ref{eq:Veff}) and the need to add higher order loop corrections. This will be considered in a separate work.

\acknowledgments

The author thanks T. Markkanen and K. Kannike for useful discussions.
This work was supported by the Estonian Research Council grants IUT23-6, PUT1026, MOBTT86  and by the ERDF Centre of Excellence project TK133.

\appendix

\section{More details about the running of $\xi$} \label{appendix}
In this Appendix we present more details about the running of the non-minimal coupling to gravity in both the metric and the Palatini formulations. We start with the metric case and reminding the corresponding tree-level action
\begin{equation}
S = \!\! \int \!\! d^4x \sqrt{-g}\left(-\frac{M_P^2}{2}f(\phi)R(\bar\Gamma) + \frac{(\partial \phi)^2}{2}  - V(\phi) \right) , \label{eq:metric:action}
\end{equation}
where
\begin{eqnarray}
f(\phi) &=& 1 + \xi \frac{\phi^2}{M_P^2} \, , \\  
V(\phi) &=& \frac{1}{4} \lambda \phi^4 \, ,
\end{eqnarray}
and $\bar\Gamma$ is the Levi-Civita connection in eq. \eqref{eq:LC}. The beta-function for $\xi$ is well known \cite{Buchbinder:1992rb}
\begin{equation}
\beta_\xi = \frac{6 \lambda}{16 \pi^2} \left( \xi+\frac{1}{6} \right) \, .
\end{equation}
Therefore it is easy to check that the running of $\xi$ is always negligible (i.e. $\frac{\beta_\xi}{\xi} \ll 1$) during the inflationary regime in the metric theory. When $\xi \ll 1$, we have
\begin{equation}
 \frac{\beta_\xi}{\xi} \approx \frac{\lambda}{16\pi^2} < \frac{1}{16\pi^2} \ll 1 \, ,
\end{equation}
while, when  $\xi \approx O(1)$ or $\xi \gg 1$, we have
\begin{equation}
 \frac{\beta_\xi}{\xi} \approx \frac{3\lambda}{8\pi^2} < \frac{3}{8\pi^2} \ll  1 \, ,
\end{equation}
where we have used in both cases the pertubativity bound $\lambda < 1$. Therefore, until the perturbativity is preserved, the running of $\xi$ can be ignored in the metric theory.

\begin{figure}[t!]
\begin{center}
 \includegraphics[width=0.75\textwidth]{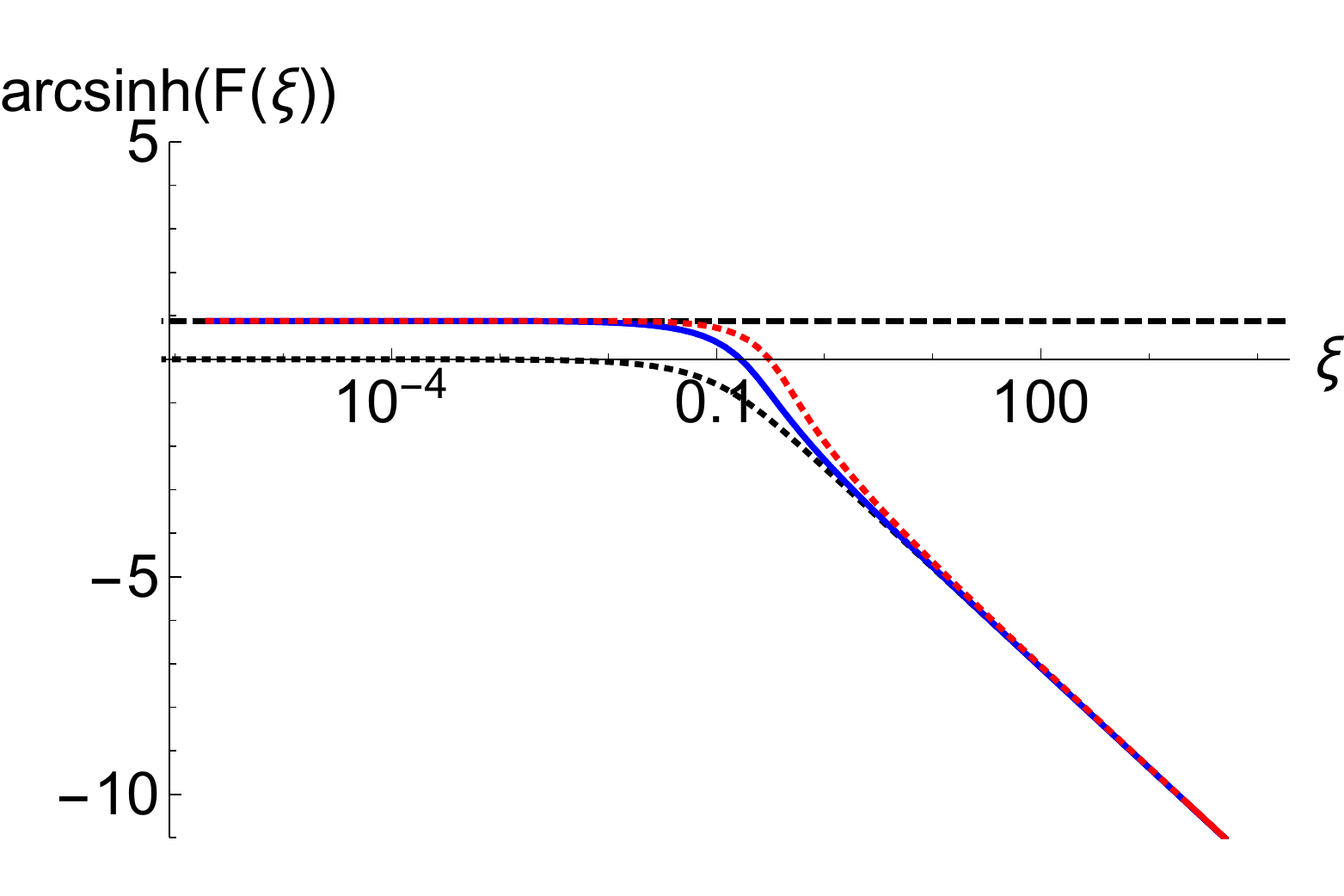}
\end{center}
 \caption{$\arcsinh(F(\xi))$ with $F(\xi)=k(\phi_f)$ (red, dotted), $F(\xi)=k(\phi_N)$ (blue, continuous), $F(\xi)=1$ (black, dashed) and $F(\xi)=-6\xi$ (black, dotted).}
  \label{Fig:non:minimal:terms:Palatini}
\end{figure}

Let us move now to the Palatini case. The corresponding action is the same as in eq. \eqref{eq:metric:action}, but with replacement of $\bar\Gamma$ with $\Gamma$ given in eq. \eqref{eq:conn:J}. However, it is possible to rewrite the Palatini action as the action of a non-minimally coupled metric theory with a non-minimal kinetic term \cite{Koivisto:2005yc}, obtaining
\begin{equation}
S = \!\! \int \!\! d^4x \sqrt{-g}\left(-\frac{M_P^2}{2}f(\phi)R(\bar\Gamma) + \frac{k(\phi)}{2}(\partial \phi)^2  - V(\phi) \right) \, , \label{eq:Palatini:metric:action}
\end{equation}
where the difference between the two formulations is now encoded in 
\begin{equation}
k(\phi) = 1- 6\frac{\xi^2 \phi^2}{\MP^2+\xi \phi^2} \, . \label{eq:k}
\end{equation}
Given the non-minimality of the kinetic term of the inflaton, an exhaustive computation of the radiative corrections (including the running of $\xi$) in the Palatini case is a task beyond the purpose of the present article. We can anyhow show a rather simple argument in support to our assumption of neglecting the contribution raising from $\beta_\xi$ also in the Palatini formulation. In order to do that, we plot in Fig. \ref{Fig:non:minimal:terms:Palatini} $\arcsinh(F(\xi))$ with $F(\xi)=k(\phi_f)$ (red, dotted), $F(\xi)=k(\phi_N)$ (blue, continuous), $F(\xi)=1$ (black, dashed) and $F(\xi)=-6\xi$ (black, dotted), where $\phi(\chi)$ 
\begin{equation}
\phi (\chi) = \frac{M_P}{\sqrt{\xi }} \sinh \left(\frac{\sqrt{\xi } \chi }{M_P}\right) \label{eq:phi:Palatini}
\end{equation}
is computed by solving and then inverting eq. \eqref{eq:dphiP},  $\phi_{f,N}=\phi(\chi_{f,N})$ and $\chi_{f, N}$ is defined after eq. \eqref{Ndef}. The choice of plotting the inverse hyperbolic sine is motivated by the need of a logarithmic scale on the $y$-axis for functions that also assume negative values. We can see that for most of the values of $\xi$ the non-minimal kinetic factors $k(\phi_{N,f})$ are either 1 or $-6 \xi$. In the first case the computation is numerically equivalent to the usual metric case (cf. Figs. \ref{Fig:Results:metric} and \ref{Fig:Results:Palatini}) and we have already shown before that the running of $\xi$ is negligible in this case. In the second case, $k(\phi_{N,f}) \simeq -6 \xi$, it is easy to check that the theory becomes conformal and therefore also in this case the running of $\xi$ is absent. To conclude we also notice that there is a small region around $\xi \approx 0.1-1$ where the $k(\phi_{N,f})$ is neither 1 nor $-6 \xi$. In this case we can only make the reasonable guess that, considering the shape and continuity of the involved functions, the running of $\xi$ still remains irrelevant. Therefore, until the perturbativity of the theory is preserved, we ignore the running of $\xi$ also in the Palatini case.

\bibliographystyle{JHEP}
\bibliography{references}

\end{document}